\documentstyle[aps,floats,epsfig,a4]{revtex}
\begin{document}
\bibliographystyle{prsty}

\title{Coherent Tunneling in Fe-8 Molecular Nanomagnets:
Response to the Comment by Bellessa (arXiv:cond-mat/0006441)}
\author{E. del Barco, J.M. Hernandez and J. Tejada}

\address{ Dept. de F\'{\i}sica Fonamental, Universitat de Barcelona, Diagonal\\
647, 08028 Barcelona, Spain}
\author{E.M. Chudnovsky}

\address{ Physics Department, CUNY Lehman
College, Bronx, NY 10468-1589, U.S.A.}
\author{E. Molins}

\address{ICMAB (CSIC), Campus Univ. Autonoma de Barcelona, 08193
Cerdanyola, Spain}

\maketitle

\vspace{1cm } All statements of Bellessa's Comment [1] are false.

To begin with, the EPL [2] that has been commented upon by Dr.
Bellessa was published not by ``Chudnovsky and Tejada" but by Del
Barco, Vernier, Hernandez, Tejada, Chudnovsky, Molins, and
Bellessa. The Fe-8 sample was synthesized in Barcelona where the
whole experiment was planned and all structural and magnetic
characterization: chemical, infrared, X-ray, magnetic, etc. (see
the EPL for details) were performed. Figure 1 shows the
magnetization curves measured in Barcelona. Only the final
measurement of the ac-susceptibility, performed by Enrique del
Barco of the University of Barcelona and Nicolas Vernier of Orsay,
used facilities of the laboratory of Dr. Bellessa. The analysis
and the interpretation of the data have been
 performed in Barcelona. The text of the paper has been agreed
upon by all authors including Dr. Bellessa, as his E-mail to Dr.
Tejada indicates [3].

\begin{figure}
\centerline{\epsfig{file=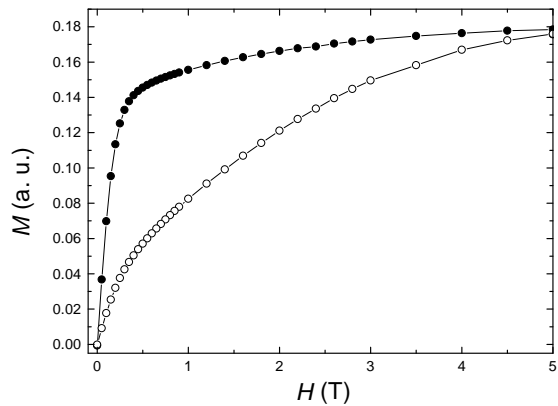,width=8truecm}}
\vspace{0.3truecm} \caption{\small Magnetization of the Fe-8
oriented powder at 2K for the magnetic field parallel (solid
circles) and perpendicular (open circles) to the orientation
axis.}
%\vspace{-0.25truecm}
\end{figure}

The question of presenting the data in terms of tunneling or EPR
is the one of semantics. Dr. Bellessa has published almost
identical data on a similar system, Mn-12, which he interpreted as
tunneling, not EPR [4]. The idea of that experiment was
``borrowed" (without our knowledge) from our EPL before it was
published.

The method of orienting superparamagnetic powder by solidifying it
in epoxy in high field is well known and has been repeatedly
tested in experiment. This is how regular steps due to resonant
spin tunneling in Mn-12 were first observed [5]. The method works
due to the anisotropy of the magnetic susceptibility of
superparamagnetic particles and has little to do with the
interaction between Fe-8 ions as is incorrectly ``suspected" by
Dr. Bellessa. Everything in his Comment about the effect of the
misalignment of the Fe-8 powder is contained in our EPL [2]. The
curves in Fig.1, which differ from the ones in the Comment,
correspond to about 35\% misalignment of easy axes, which is
stated in our EPL. Fig. 1 of the Comment was measured by Dr.
Bellessa many months later and (assuming the measurement was
correct) must be attributed to the well-known aging of the
oriented powder sample.

\end{document}